\documentclass[11pt,onecolumn]{article}    
\usepackage{abstract}
\usepackage{authblk}
\usepackage{graphicx}
\usepackage{xcolor}
\usepackage{tabularx}
\usepackage[normalem]{ulem}
\usepackage{amsmath}
\usepackage[utf8]{inputenc}
\usepackage{natbib}
\usepackage{afterpage}
\usepackage[bindingoffset=0.2in,%
            left=1in,right=1in,top=1in,bottom=1in,%
            footskip=.25in]{geometry}
\usepackage{todonotes}
\usepackage{listings}

\lstdefinelanguage{bngl}{
	alsoletter={:,-,<,>,@,+},
	morekeywords=[1]{->, <->},
	morekeywords=[2]{\:,\+,\.}
	morestring=[d]{'},
	morecomment=[l]\#
}
\begin{document}

\title{Generalizing Gillespie's direct method to enable network-free simulations}

\author[1,2]{Ryan Suderman}
\author[1]{ Eshan D. Mitra}
\author[1,2]{Yen Ting Lin }
\author[1]{Keesha E. Erickson}
\author[1,2]{Song Feng}
\author[1,2,*]{William S. Hlavacek}

\affil[1]{Theoretical Division and Biophysics Group, Los Alamos National Laboratory, Los Alamos, New Mexico 87545, USA}	
\affil[2]{Center for Nonlinear Studies, Los Alamos National Laboratory, Los Alamos, New Mexico 87545, USA}	
\affil[*]{Corresponding author: wish@lanl.gov}

	\maketitle
    \begin{abstract}
{Gillespie's direct method for stochastic simulation of chemical kinetics is a staple of computational systems biology research.  However, the algorithm requires explicit enumeration of all reactions and all chemical species that may arise in the system.  In many cases, this is not feasible due to the combinatorial explosion of reactions and species in biological networks.  Rule-based modeling frameworks provide a way to exactly represent networks containing such combinatorial complexity, and generalizations of Gillespie's direct method have been developed as simulation engines for rule-based modeling languages.  Here, we provide both a high-level description of the algorithms underlying the simulation engines, termed network-free simulation algorithms, and how they have been applied in systems biology research.  We also define a generic rule-based modeling framework and describe a number of technical details required for adapting Gillespie's direct method for network-free simulation.  Finally, we briefly discuss potential avenues for advancing network-free simulation and the role they continue to play in modeling dynamical systems in biology.}
\end{abstract}
\vspace{10pt}

\section{Introduction} \label{sec:intro}

In a living cell, numerous biochemical species interact with each other, forming complex reaction networks.
How the cell functions is largely determined by the dynamics of these networks. 
One of the goals of systems biology is to understand the emergence of phenotypes from the complex interactions present in these reaction networks.

Mathematical modeling and simulation are powerful tools for studying biochemical reaction networks.  
The interactions between chemical species can be rigorously defined and simulated using diverse techniques for representing nonlinear dynamical systems.
Exploratory analyses and hypothesis testing can be performed efficiently in a computational setting.
As a result, computational systems biology has been thriving over the past decade and has become a recognized field within quantitative biology.  

\subsection{Modeling chemical reaction networks}

Traditionally, reaction networks have been modeled with systems of ordinary differential equations (ODEs) that are solved with numerical integration algorithms.  
These models are typically constructed on the basis of mass action kinetics. 
In many cases, this approach is appropriate and successful \citep{le2006biomodels,Deuflhard2015}.
However, the ODEs describe the well-mixed concentrations of chemical species and may be inappropriate in the context of a cell because chemical species may be present in low copy numbers. 
A typical eukaryotic cell has a small volume, on the order of picoliters, and the enclosed chemical species have finite populations.
Consequently, intrinsic noise \citep{Elowitz2002,kepler2001stochasticity,ozbudak2002regulation,blake2003noise,thattai2004stochastic,kaern2005stochasticity,acar2008stochastic,munsky2009listening,lin2016gene,lin2016bursting,hufton2016intrinsic,lin2017efficient,lin2017stochastic} due to the finite and discrete nature of the reactants could be an important factor to consider when studying the dynamics of intracellular reaction networks.

Stochastic and discrete-state models provide a sensible solution to describe the fundamentally stochastic biochemical reactions between finite and discrete reactants. 
Among them, continuous-time Markov chains (CTMC) have become a standard way to represent and simulate biochemical reaction networks. 
Formally, the joint probability distribution of a CTMC is described by a Chemical Master Equation (CME).  
Except for special cases, it is impossible to derive the full analytical solution and efficient numerical techniques must be employed to solve the CMEs. 
There are two ways to numerically solve the CME: the first way is to directly integrate or approximate the CMEs \citep{Munsky2006,Cao2016}, and the second way is to use various continuous-time Monte Carlo techniques to generate sample paths of the random processes and use the sample paths to compute statistical quantities of interests \citep{bortz1975new,gillespie1976general,gillespie1977exact}.
In this review, we focus on the second approach.

\newcommand{\textoperatorname}[1]{%
  \operatorname{\textnormal{#1}}%
}

Using Monte Carlo approaches to sample trajectories from dynamical systems can be traced back to the late 1940's in Los Alamos where Enrico Fermi and Robert D. Richtmyer \citep{FermiRichtmyer1948} and Nicolas Metropolis and Stanis\l{}aw M.~Ulam \citep{metropolis1949monte} proposed the independent random sampling idea to solve problems in kinetic theory (Boltzmann and Fokker--Planck equations).
While a dynamical interpretation was clearly given in \citep{metropolis1949monte}, the proposal soon evolved to the famous Metropolis algorithm \citep{metropolis1953equation}, which did not capture time dependence and focused instead on the equilibrium distribution of a thermal system.
In the 1960's, several seminal papers generalized the idea of Monte Carlo sampling to dynamical processes in continuous time \citep{coxMiller1965theory,young1966monte}. 
Soon, the technique was applied to various problems from material science to statistical physics (a review of topics can be found in \citep{Voter2007}).
The first rejection-free algorithm, the n-fold BKL algorithm, was proposed by Bortz, Kalos, and Lebowitz in 1975 \citep{bortz1975new}.
With this algorithm, the advanced time is a function of all the possible transition events which may not take place in a specific sample path. 
Soon, Gillespie introduced such rejection-free kinetic Monte Carlo methods to the study of chemical reaction networks \citep{gillespie1976general}.
The method became a standard way to simulate such networks and is often termed Gillespie's algorithm or the stochastic simulation algorithm (SSA).
Ever since Gillespie's paper, there have been many proposals to improve the performance of the SSA by introducing novel data structures, such as indexed priority queues \citep{gibson2000efficient} or constructing an implementation that takes advantage of some general structure of the reaction network, such as a sparse transition matrix \citep{Ramaswamy2010}; however, the core of the algorithm remains intact.

An intrinsic, often unstated, assumption of the SSA is that the chemical reaction network must be fully specified. 
However, in biology we commonly only have data about pairwise interactions.
Further complicating the situation is the fact that biochemical reaction networks often involve macromolecules (e.g., proteins) that have multiple domains for interacting with other molecules and that may each occupy a number of chemical states.  For example, a protein may have multiple residues that are subject to phosphorylation or dephosphorylation.  
The platelet-derived growth factor receptor (PDGFR) has an intracellular domain containing 10 amino acid residues that may be phosphorylated.  If each residue is independently subject to phosphorylation and dephosphorylation, a single PDGFR molecule can occupy $2^{10} = 1024$ possible biochemical states \citep{Mayer2009}.  However, the PDGFR dimerizes upon binding a ligand, increasing biochemical state space to over 500,000 possible states \citep{Mayer2009}.  Representing the PDGFR as a system of differential equations would thus require over 500,000 equations.  
Similarly, biochemical species can form large complexes through pairwise binding, and in some cases the total number of possible biochemical species exponentially increases with the number of distinct proteins in the model \citep{Suderman2013,Faeder2005,Bray2003,Endy2001}.
This vastness in state space is termed combinatorial complexity.

\subsection{Representing complex systems simply}
Rule-based modeling was introduced in systems biology to address the problem of combinatorial complexity, particularly for models of intracellular cell signaling networks \citep{Chylek:2014io,Danos2007rule}. Its core insight is the use of rules to represent a class of reactions that operate on identical reaction centers.  Specifically, rules use molecular patterns common among multiple reactant molecules, allowing for more succinct representation of reactions, which precludes the need to enumerate all possible reactions between all possible biochemical species \citep{Chylek:2014io}.  Rule-based modeling enables a precise and comprehensive, yet tractable, representation of complex systems \citep{deOliveira2016}. With a rule-based approach, a modeler can represent the previously described $>$500,000-state PDGFR system with a mere 22 rules, assuming mutual independence of the rules involved (Code \ref{lst:pdgfr}): 2 ligand-receptor interaction rules (binding and unbinding), 2 receptor dimerization rules (binding and unbinding), 10 phosphorylation rules and 10 dephosphorylation rules (one of each per residue).  The complete formulation of this model, written in the BioNetGen language \citep{Faeder2009}, is given in Appendix \ref{sec:pdgfr}.

Rule-based modeling in systems biology was initially developed to automate construction of biochemical reaction networks, as describing biological phenomena with rules significantly reduces the work of model construction \citep{Blinov2006}. However, the resulting network can be of intractable size from a computational perspective, or it may even be unbounded except by the number of molecules present in the system \citep{Yang2008kinetic}. A natural generalization to mitigate this issue is to generate the network on-the-fly \citep{Lok2005automatic,Faeder2005rule,Faulon2001}. Even if the number of chemical species can be (possibly infinitely) large, on-the-fly network generation assumes that only a finite (and sometimes small) number of species are typically populated. Therefore, instead of insisting on calculating reaction rates for all possible reactions, which may be impossible, one generates the local network (and thus determines reaction rates) for the reactions whose reactants are present in the system.  As the system evolves, this local network is updated when new species are introduced or existing species are removed completely.  While sometimes useful, the size of the boundary (the part of the network requiring generation of new reactions) may increase exponentially, and so simulation becomes very inefficient in such a case.

This motivates an approach that avoids generating the chemical reaction network altogether: instead, each molecule is treated as an object and the simulation is performed directly on the objects. This \emph{network-free} approach, is a form of agent-based simulation grounded in physical and chemical principles. Various network-free software packages have been deployed throughout the years, most of which correspond to a particular rule-based modeling framework  \citep{Yang2008kinetic,Colvin2010RuleMonkey,LeNovere2001STOCHSIM,Colvin2009simulation,Danos2007scalable,Sneddon2011efficient,Zhang2005,Sweeney2008}. 
The algorithm implemented in these simulators is a modified form of Gillespie's direct method \citep{gillespie1976general,gillespie1977exact}.  The modified algorithm differs from Gillespie's method in that careful bookkeeping of the state of the system is required.

Throughout the rest of this work, we provide a brief review of research applications of existing implementations of network-free simulation algorithms (Section \ref{sec:applications}) followed by a high-level description of how Gillespie's direct method can be generalized for a rule-based modeling framework (Section \ref{sec:genericAlgo}). We then define a generic rule-based modeling framework and proceed to describe a basic implementation of a network-free simulation algorithm (Sections \ref{sec:nomen} and \ref{sec:distinct}). Finally, we investigate a number of technical details that need consideration when implementing such an algorithm (Sections \ref{sec:complexity} and \ref{sec:conventions}).

\section{Applications of network-free simulation}\label{sec:applications}
A number of software packages exist to facilitate network-free simulation for specific rule-based modeling languages. Two of the most prominent rule-based modeling languages for biomolecular simulation are the Kappa language \citep{Danos2004formal} and the BioNetGen language (BNGL) \citep{Faeder2009}. These languages each have associated network-free simulation tools: KaSim for the Kappa language \citep{KaSim} and NFsim \citep{Sneddon2011efficient} (among others) for BNGL. Both the KaSim and NFsim engines have been used in situations where model complexity necessitates network-free simulation.  

A straightforward application of network-free simulation seeks to understand how combinatorial complexity influences the protein complexes that assemble during signal transduction.  Explicitly investigating combinatorial complexity precludes any notion of species or reaction network enumeration, and so network-free simulation is the only path forward.  Deeds et al. \citep{Deeds2012} constructed a rule-based model in the Kappa language that characterized the protein-protein interaction network in yeast. They found that existing knowledge of the yeast interaction network led to extreme combinatorial complexity. Ultimately, they found that simulations beginning from the same initial state diverged rapidly in the sets of complexes that were formed, where any two independent simulations exhibited only 20\% overlap among unique complexes formed \citep{Deeds2012}. The same methodology was applied to the pheromone signaling network in yeast, this time revealing that reliable signal transduction can occur in networks even when the assembled molecular complexes responsible for signaling are highly variable \citep{Suderman2013}. These bodies of work highlight two key facts about complexity in interaction networks: first, that clonal cells are likely never in the same state at the same time, and second, that cells can still reliably process extracellular information despite a seemingly chaotic environment. 

Intracellular signaling in mammalian cells often involves oligomerization or polymerization of cell surface receptors \citep{Su2016}. These receptors also recruit a wide array of downstream effector proteins that govern the dynamics of signal transduction.  Examples of these systems include growth factor signaling \citep{Creamer2012,Stites2015} and the antigen recognition receptor signaling \citep{Chylek2014,Nag2010,Nag2009}. Because of the formation of complex multimeric structures, the state space of these intracellular signaling models is bounded only by the number of molecules in the simulation. Using network-free simulation algorithms is often the only approach available to characterize system dynamics without undue simplification of a model. Most of the work on mammalian signaling referenced here used models written in BNGL and were simulated with a BNGL-compatible network-free simulator, NFsim.  

Another domain of research involving biopolymers considers nucleic acids, and one example of nucleic acid research facilitated by rule-based modeling examined the phosphorylation states of RNA Polymerase II as it binds to an arbitrary number of positions on a DNA sequence \citep{Aitken2013}. In this case, the number of possible biochemical states is susceptible to the problem of combinatorial complexity as the size of the DNA molecule increases. A second example focused on base excision repair in DNA \citep{Kohler2014}. In this work, the model includes a single DNA molecule composed of 100,000 base pairs, something that would be utterly intractable in any other modeling framework. The model incorporates a number of protein-protein interactions in addition to a mechanistic representation of base repair catalysis involving endonucleases, polymerases, and ligases to investigate how certain molecules (such as scaffold proteins) contribute to the speed and efficacy of DNA repair.  

Finally, rule-based modeling also has applications outside biology. One example explores the usefulness of applying rule-based modeling and network-free simulation in simulating labor markets \citep{Kuhn2016}. In particular, it highlights the differences between general agent-based modeling frameworks and rule-based modeling, which is a type of agent-based modeling that is based on formal chemical kinetics \citep{Danos2007scalable,Faeder2009}. A notable difference between existing agent-based modeling frameworks for modeling labor markets and rule-based modeling coupled to network-free simulation is the absence of spatial information in network-free simulation (i.e., the well-mixed assumption). Although these examples serve as a reminder for the general applicability of rule-based modeling approaches and network-free simulation algorithms, we focus primarily on biological applications for a detailed description of the network-free simulation approach.

\section{A minimalist description of network-free simulations} \label{sec:genericAlgo}

\begin{figure}[t]
\centering
\includegraphics[width=0.65\textwidth]{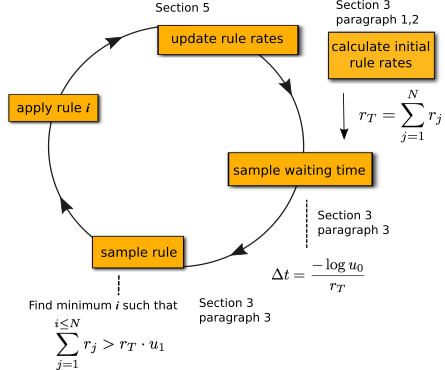}
\caption{A simple graphical depiction of a generalization of Gillespie's direct method, where $N$ is the total number of rules. Sections describing the relevant steps are labeled in the figure. After initially computing the rates for every rule, each iteration of the simulation loop requires generating two pseudorandom numbers $u_0$ and $u_1$ on the interval [0,1). The waiting time until the next event, $\Delta t$, is calculated from the first random number and it is inversely proportional to the total rate of the rules in the system: $r_T$. Then rule $i$ is sampled, where $i$ is the smallest number such that the cumulative rate of rules 0 through $i$ is greater than the product of the second random number and the total rate, $r_T$.  After both the next event time and rule (i.e., the type of reaction) are sampled, the rule is applied to a specific set of molecules from among all molecules that qualify as reactants, and the rates are updated based on the change of the system.}
\label{fig:loop}
\end{figure}

The algorithm to generate exact sample paths for a rule-based model using a network-free approach is largely similar to Gillespie's direct method as seen in Fig. \ref{fig:loop} \citep{gillespie1976general,gillespie1977exact}.
In this section, we focus on a high-level general description of the algorithm, leaving the technical (but important) details specific to network-free simulations to the next sections.  This description includes some jargon that we define more rigorously in Section \ref{sec:nomen}: 
\begin{itemize}
\item a molecule is an explicit object that is tracked by the simulation engine
\item a rule is composed of patterns that identify molecular moieties in reactant molecules
\item a pattern is used to match reactant molecules during simulation
\item a rule defines a chemical transformation that is to be applied to a reactant molecule
\item a species defines a class of some molecular complex (i.e. a particular biomolecular configuration)
\item a mixture explicitly defines all interacting molecules in the simulation
\end{itemize}
We decompose the algorithm into four essential blocks: initialization of the system, computation of the rules' rates, advancing time and sampling a rule, and updating the system.

\paragraph{Step 1. Initialization.} At the beginning of the simulation, the system configuration is defined by the user. Individual instances of all the chemical species' component molecules are populated.  

\paragraph{Step 2. Computation of rule rates.} In this step, matches are constructed between the patterns defined in the rules and the explicit molecule instances in the simulation mixture.  One of the major differences between rule-based models and traditional models is that chemical transformations are defined in terms of patterns instead of in terms of chemical species \citep{gillespie1976general,gillespie1977exact}.  This means that many distinct chemical species may participate as a reactant in a rule, provided each of these species matches the same reactant pattern of the rule.  The rate of the rule is then proportional to the number of matches of its reactant patterns.

\paragraph{Step 3. Advancing time and sampling a rule with specific reactants.} Once all the rules' rates are computed, the time to the next rule application is a random number sampled from an exponential distribution whose rate parameter is the sum of all the rules' rates: $r_T$.  A random waiting time $\Delta t$ is sampled from this exponential distribution
\begin{equation*}
\Delta t = \dfrac{-\log{u_0}}{r_T}
\end{equation*}
where $u_0$ is a uniform random number on the interval [0,1).  The system's simulation time is then advanced by $\Delta t$.  The rule to be applied to the system is sampled with probability proportional to its rate.  This is done via the conditioning procedure: 
\begin{itemize}
\item Sample a uniform random number $u_1$ on the interval [0,1).  
\item Iterate over all rules to find the minimum rule index, $i$, such that 
\begin{equation*}
\sum^{i\le N}_{j=1} r_j > r_T \cdot u_1
\end{equation*}
where $N$ is the number of rules.
\end{itemize}
These steps are identical to Gillespie's direct method \citep{gillespie1976general,gillespie1977exact}. However, since the rules are defined by patterns, the sampled rule does not specify which molecule instances should be modified by the transformation defined by the rule. The simulator therefore samples molecules with uniform probability from a list of matches for each of the selected rule's reactant patterns.

\paragraph{Step 4. Updating the system's configuration.} After the rule is applied (modifying the matched molecules sampled in Step 3) the matches between rule patterns and the molecules in the simulation mixture are updated.  To increase simulation efficiency, the rules' rates can be updated incrementally to avoid recalculating all the matches.  After the rules' rates are updated, the simulation continues from Step 3 until a stopping condition is met.  We provide a detailed discussion of how this step can be implemented in Section \ref{sec:distinct}.\\

Although the above description is very similar to Gillespie's original direct method \citep{gillespie1976general,gillespie1977exact}, interfacing the algorithm with a particular rule-based modeling framework requires careful consideration of how pattern-matching affects a rule's rate.  We discuss a number of relevant issues that may arise in building a network-free algorithm in Sections \ref{sec:complexity} and \ref{sec:conventions}.

\section{Nomenclature}\label{sec:nomen}
In this and the following sections, we consider a number of technical details required for implementing a network-free simulation algorithm. To do so, we first must define the objects involved in the simulation. Different rule-based modeling frameworks use distinct nomenclature for similar, sometimes identical, constructs. Additionally, many descriptions of rule-based modeling approaches contain jargon that can easily confuse readers without the relevant domain-specific knowledge\footnote{Most objects in rule-based modeling can be represented visually (and formally) as graphs. Rules then become transformations (i.e., rewriting operations) on these graphs, and much of the jargon relating to rule-based modeling has its origin in graph theory. We will occasionally mention these terms, but will not rely on them.}. Here, we try to provide intuitive and sufficiently precise definitions of the objects required to construct a network-free simulation algorithm independent of any specific rule-based modeling framework. Note that our terminology reflects the usage of rule-based modeling in molecular and cellular biology \citep{Chylek:2014io,Danos2008}.  

\afterpage{
\begin{figure}[t]
\centering
  \includegraphics[width=0.75\textwidth]{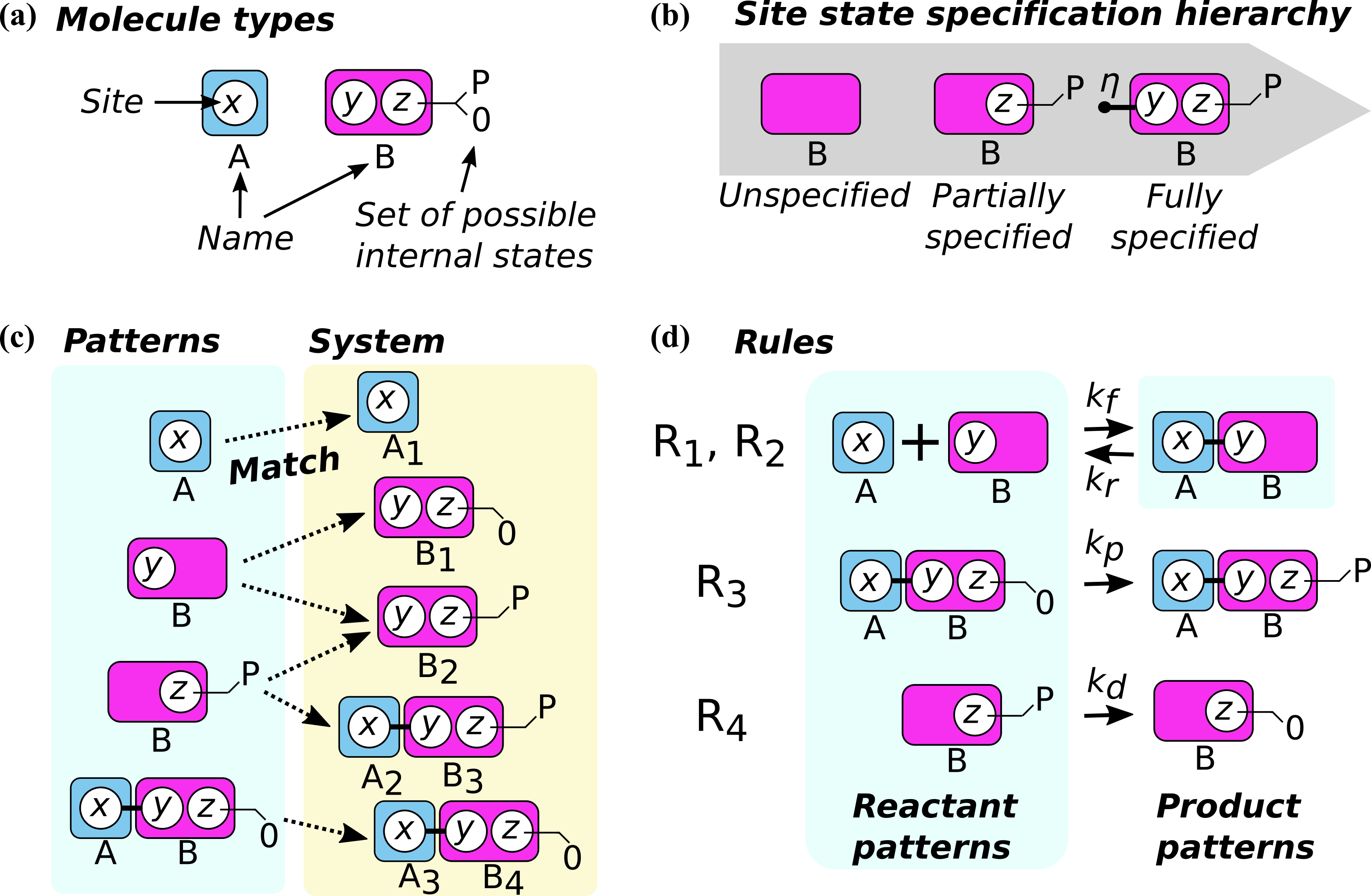}
\caption{Nomenclature for rule-based modeling. {\bf a} Molecule types are archetypes for individual objects (molecules) in a simulation. Each molecule type has a name and a list of sites. Sites contain information about binding state, and may occupy a state from a set of possible internal states. {\bf b} Molecules can be ordered based on their degree of specification.  Note that such an ordering assumes that information present in less specific molecules is preserved in more specific molecules.  A molecule may have no specified sites.  A partially specified molecule includes some information about binding site state (site \emph{z} is phosphorylated). For full specification, the state of every site is explicit (site \emph{y} is bound with edge \(\eta \) and site \emph{z} is phosphorylated. {\bf c} A match exists when a pattern's sites are either equivalent to or less specific and consistent with the site states in corresponding molecules in a simulation mixture (the subscript on the molecules are unique integer identifiers for molecules of a particular type). {\bf d}  Four rules are defined (R$_1$ to R$_4$): association/dissociation of molecules A and B, phosphorylation of B when bound to A, and dephosphorylation of B.  A rule consists of reactant patterns, a transformation defined by product patterns, a rate law, and (optionally) any other conditions that influence the rule.  Typically reactants are written on the left-hand side of the rule and products are written on the right hand side.  However, note that since the first rule is reversible (i.e., it has the double-arrow operator), the right-hand side serves as the reactant pattern for the dissociation reaction and the left-hand side contains the product patterns.}
\label{fig:1}       
\end{figure}
\clearpage
}
The atomic unit of rule-based modeling applied to biochemical reactions is what we term the \emph{molecule}. A molecule is typically a representation of a biological macromolecule, such as a protein, but a molecule could also represent a metabolite, drug, other small molecule, or any object of interest in the model.  Rule-based modeling frameworks often require molecules to be first defined with a particular signature, and specific instances of these molecules (such as the molecules tracked during simulation) must conform to this definition.  Such signatures are composed of a name that acts as a label for the molecule's type, as well as a predefined list of \emph{sites} that typically represent physical or chemical attributes of the molecule (Fig. \ref{fig:1}a). Sites are defined with names\footnote{To our best knowledge, BioNetGen is the only framework that allows molecules with multiple identically-named sites.  These sites are treated as equivalent.  Molecule types must have unique names.}, and a predefined list (which can be empty) of internal states that represent any other property of the site (e.g., whether an amino acid residue in a protein is phosphorylated) (Fig. \ref{fig:1}a).   Sites also engage in binding to other sites, allowing association between molecules and thus representation of higher order molecular structures.  Modification of sites' states (both internal and binding) typically comprise the majority of rule applications (reaction events) during simulation (e.g., binding, unbinding, or chemical modification).\footnote{Rules may also define synthesis and degradation of molecules.}  

Molecules that are used for representing specific molecular moieties and not specific instances of physical objects are termed \emph{pattern molecules}.  Sites in pattern molecules may be completely absent (unspecified) or a site's states may be partially specified to capture the necessary and sufficient features required for representing a molecular moiety.  Of course, partial specification of a site's states (binding or internal) relies on syntax to convey incomplete knowledge about the site's states.  For example, the modeling language used to write a rule may be able to express whether a site is bound, but its binding partner is unknown, or whether the site's internal state is in some subset of its predefined set of internal states.  Furthermore, this partial specification has an ordering, where a site's absence or partial specification of its states in a pattern molecule is less specific than a site with a more specific or complete specification of its state (Fig. \ref{fig:1}b).  The notion of such an ordering is useful in pattern-matching; a less-specific object may match a more-specific or completely-specified object provided that the information in the less-specific object is preserved in the more-specific object.  This concept of preservation of information is implicitly assumed in later discussions of partial specification of molecules in pattern matching.  Similarly, if the internal state of one site is identical to another site and the binding states for the sites are both explicitly bound or explicitly free, then we say that the sites are equivalent.  Sets of sites (and thus molecules) can similarly be ordered according to specificity or equivalency.   

We define \emph{complete molecules} or simply \emph{molecules}, as molecules whose sites are all fully specified.  Full specification requires that sites are either bound to another site with a labeled bond\footnote{A labeled bond links two sites, meaning that both partners in a bond can be determined by the bond label.} or unbound, and that sites express an explicit internal state assuming that they have predefined internal states to occupy.  Complete molecules are those that are involved in the simulation, being modified by rule applications.

We can now define larger objects composed of complete molecules and pattern molecues.  An object composed of a list of one or more complete molecules that only connect to other molecules in their list (a \emph{connected component} in graph terminology) is called a \emph{species}, reflecting standard chemical nomenclature\footnote{In chemistry and ecology, the term \emph{species} refers to a class of things (molecular configurations or organisms).  We retain this convention and refer to specific \emph{instances} of a species when discussing an individual object that conforms to the features that define a particular species.}.  Note that a molecule with no bound sites is both a molecule and a species.  Similarly, we define a \emph{pattern} as a list of pattern molecules that are explicitly connected only amongst themselves (Fig. \ref{fig:1}c, blue region elements)\footnote{In some cases, patterns may be defined as involving unconnected molecules, but we do not adopt this convention}.  Finally we use the term \emph{mixture} or \emph{simulation mixture} to refer to the pool of complete molecules that are interacting during the simulation of a rule-based model (Fig. \ref{fig:1}c, entire yellow region).  Network-free simulation engines track individual objects in a mixture, as opposed to Gillespie's direct method, which tracks the populations of predefined chemical species.  Note that populations of chemical species can be reconstructed from the list of complete molecules that defines a mixture, and so ours is an equivalent representation of the state of the system compared to the traditional SSA for the purpose of tracking population dynamics.  

To avoid the need to a priori generate the reaction network, rates are calculated via pattern matching.  We define a \emph{match} as a mapping from a pattern to a list of molecules in a simulation mixture (Fig. \ref{fig:1}c).  This association depends on two criteria:
\begin{enumerate}
\item Each pattern molecule in the pattern must have a corresponding molecule of the same name in the mixture\\
\item For a pattern molecule $X$ in the pattern and its corresponding molecule $Y$ in the mixture:\\
\begin{enumerate}
\item if $X$'s sites are less specific than $Y$'s sites, then the information present in $X$'s sites must be preserved in $Y$'s sites (consistency)\\
\item sites with fully specified states present in $X$ must have equivalent corresponding sites in $Y$
\end{enumerate}
\end{enumerate}

A \emph{rule} defines a set of reactions and involves all of the previously defined concepts (Fig. \ref{fig:1}d).  A rule is composed of a list of reactant patterns and a list of product patterns that together define a transformation (Fig. \ref{fig:1}d, blue regions), as well as a rate law.  The reactant and product patterns are commonly known as the \emph{left-hand side} and \emph{right-hand side} of a rule, respectively.  The rate law is used to calculate a rule's \emph{rate}, which is the sum of all of the rates of all of the reactions implied by the rule.  Each reaction implied by a rule inherits the rate law associated with the rule.

Finally, we use the term \emph{system} to generally describe everything needed to fully represent and simulate the model.

\section{Distinctions from traditional SSA}\label{sec:distinct}

The generalization presented in Section \ref{sec:genericAlgo} differs from a conventional SSA in a few ways. Two early examples of the general approach described here can be seen in \citep{Yang2008kinetic} and \citep{Danos2007scalable}. Rates are dependent on numbers of matches instead of population sizes of species. Another related distinction is the need to sample molecules that will be altered by a rule once the rule has been selected for application. Perhaps the most notable and complicating difference is the methodology used to store and update the state of the system.  This section provides a high-level view of the data structures necessary to accommodate these differences in a reasonable manner. We make no claims regarding the efficiency of this approach, but present it as a means to understand the essential steps of constructing a network-free simulation algorithm.

\subsection{Computing the rule rates}
To calculate the initial rates of the rules in a model, we first count the number of matches associated with each reactant pattern in a rule. A simple way to store this information is to have, for each reactant pattern in each rule, a \emph{match list} containing all of the matches from a pattern to the molecules in the simulation mixture (Fig. \ref{fig:alg}a). To initialize the match list for each pattern, we select an arbitrary pattern molecule in the pattern, which we will refer to as the anchoring pattern molecule. We then find all molecules in the initial mixture that match the anchoring pattern molecule: the anchoring molecule.  We recursively traverse the species instance that contains the anchoring molecule (which is possible because the molecules contain pointers referencing the other molecules to which they are bound) to determine all the unique matches arising from the choice of anchoring molecule\footnote{The anchoring molecule simply serves as an arbitrary starting point for traversing the species instance (i.e. multiple starting points may be possible).}. The set of matches forms the initial match list of the pattern.  For rules with elementary rate laws (i.e., rate laws consistent with mass-action kinetics), the rate will typically be the rule's rate constant multiplied by the product of the numbers of matches for all reactant patterns.  A simple example of this can demonstrated for rule $R_1$ (Fig. \ref{fig:1}) from the information present in Fig. \ref{fig:alg}a:  
\begin{enumerate}
\item Determine the number of matches for each reactant pattern
\begin{itemize}
\item There are 2 matches for rule $R_1$, pattern $P_1$
\item There are 2 matches for rule $R_1$, pattern $P_2$
\end{itemize}
\item Calculate the product of the numbers of matches: $2\cdot2=4$
\item Calculate the rate by multiplying the product of matches by the rule's rate constant: $k_f\cdot4$
\end{enumerate}
However, there are a few issues to consider that may affect the calculation, and these issues are discussed in Section \ref{sec:complexity}.

\subsection{Sampling the reactants}
Provided that a match list is correctly maintained for each reactant pattern in each rule, the sampling step of the simulation is straightforward (Fig. \ref{fig:alg}b).  A rule is sampled according to the method described in Step 3 of Section \ref{sec:genericAlgo}.  After selection of a rule, it is necessary to sample which molecules will be modified by the transformation defined by the rule.  To make this choice, a match for each pattern in the rule is sampled with uniform probability from the corresponding match lists, and the rule is applied to the reactant molecules in the mixture that correspond to the selected match.  

\subsection{Updating the system}
When a reaction event\footnote{Nonproductive (null) events are described in Section \ref{sec:null}.} occurs, the state of the system must be updated appropriately (Fig. \ref{fig:alg}c). If a transformation changes a site's internal state, that state must be updated on the molecule, and if the transformation includes the addition or removal of bonds, the molecules involved must have the pointers to their binding partners updated accordingly. Finally, the most computationally expensive step is to update the match lists.  A brute force approach is to check all of the reactant patterns of all rules to determine if their match lists have been affected by the transformation that has just been applied.\footnote{Depending on how patterns are specified, a brute force approach may be the only way to correctly update the system. See Section \ref{sec:local}.}

After a reaction event, the first phase of the update process, termed the negative update phase (in accordance with the terminology of \citep{Danos2007scalable}), is to consider the molecules that were altered or removed by the reaction event. In this phase, we traverse the species instances containing the molecules that were involved in the reaction before the rule application\footnote{This traversal, and that in the positive update phase, allows the algorithm to accommodate rules with nonlocal constraints. See Section \ref{sec:local}.}, and remove the matches involving those species instances from all match lists.

The second phase, termed the positive update phase (in accordance with the terminology of \citep{Danos2007scalable}), is to consider the newly created species instances formed by the reaction event. Similar to the negative update phase, the new species instances must be fully traversed to determine which rules' patterns need to be updated.  Any new matches are then added to the appropriate match lists. 

After the negative and positive update phases, the match lists have been fully updated, and the changes to the match lists can be used to efficiently update rule rates.  Rate updates are most efficient if current rates are modified up or down in accordance with match list changes, as opposed to de novo rate calculation (i.e., calculating the rate by constructing the match lists from scratch).
\afterpage{
    \begin{figure}[t]
\centering
  \includegraphics[width=0.65\textwidth]{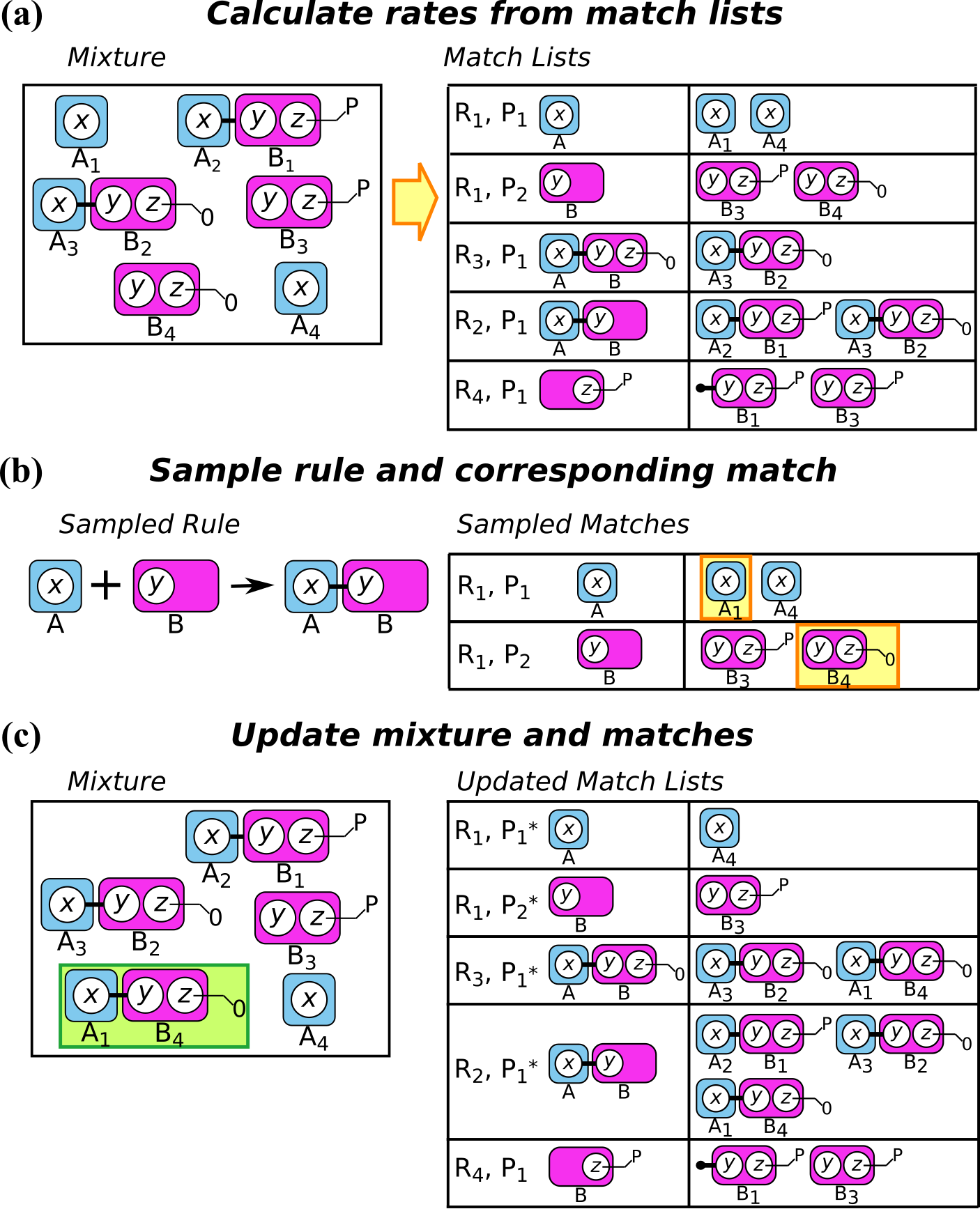}
\caption{Schematic of a network-free simulation algorithm. {\bf a} At any point during a simulation, the mixture is stored in memory, as is the match list for each pattern.  Note that $R_\text{N}$, $P_\text{M}$ refers to the N$^\text{th}$ rule and M$^\text{th}$ pattern in that rule. The rules considered here are those presented in Fig. \ref{fig:1}.  Rates for each rule can be calculated based on the rule's rate law and its match lists. {\bf b} At each simulation step, a rule is chosen with a probability given by its rate relative to the overall rate of all rules. A match is then chosen from the match list of each reactant pattern in the rule. {\bf c} The simulation mixture is updated according to the transformation defined in the rule.  The species highlighted in green is the result of the transformation in panel b.  The match lists are then updated; for each modified molecule, matches that have become invalid are removed, and new matches are added to the appropriate match lists.  Match lists with an asterisk have been modified from the initial system state in panel a. }
\label{fig:alg}       
\end{figure}
\clearpage
}

\section{Considerations for rate calculation} \label{sec:complexity}

A number of issues arise as a result of using rules and a pattern matching algorithm instead of the traditional SSA's use of species' populations and reactions.  Here, we discuss some of the more prominent issues relevant for accurate and consistent calculation of rule rates.   

\subsection{Symmetry among patterns}
In cases where there is symmetry among a rule's reactant patterns (i.e., when a rule's reactant patterns contain two or more identical patterns), one must take care to correctly count the number of reactions that may occur.  In Gillespie's original notation \citep{gillespie1976general}, $h_\mu$ is the number of ways a reaction may occur, and it is computed from the populations of the chemical species involved as reactants in the reaction.  The product of $h_\mu$ and the reaction's rate constant is the reaction's rate.  With our proposed method for tracking matches, we must similarly count the number of distinct match combinations in the presence of symmetry:
\begin{equation*}
h_{sym, T} = \binom{|T|}{N_T}
\end{equation*}
where $N_T$ is the number of times pattern $T$ is present in the list of reactant patterns, and $|*|$ denotes the size of $T$'s match list.  A simple example is a homodimerization rule with a rate constant $k$ for some molecule $A$.  If we consider a mixture with 1000 monomeric $A$ molecules, then we can calculate the rule's rate:
\begin{equation*}
\binom{1000}{2}\cdot k = \dfrac{1000\cdot 999}{2} \cdot k
\end{equation*}

\subsection{Symmetry within a pattern}\label{sec:permute}
Patterns may be defined in such a way that multiple matches may arise from a simple permutation of the molecules involved in the match.  For example, a pattern may involve two identical pattern molecules that are bound to each other on identically named sites.  In such a case, the pattern may match the same set of molecules more than once.  In Fig. \ref{fig:complex}a, an example is shown for a simple dimer dissociation reaction.  A single dimer exists in the simulation mixture, and the reactant pattern matches the dimer such that the first pattern molecule matches molecule $A_1$ and the second pattern molecule matches molecule $A_2$ (blue).  By permuting the molecules in the mixture (substituting $A_1$ for $A_2$ and vice versa), we find a second match between the reactant pattern and the dimer (red).  Correctly calculating the rule's rate requires dividing the rate by the number of molecule permutations in the pattern that preserve bond connectivity.\footnote{When representing patterns as graphs, this number is the order of the \emph{automorphism group} of the graph, roughly the number of ways a graph can be mapped to itself.}  The divisor similarly corresponds to the number of ways an occurrence of the pattern in a species instance is matched by the pattern because of symmetry in the pattern.  This rate calculation assumes a specific convention about the semantics of our framework: a rule's rate should be proportional to the number of distinct reactions that can occur, and not simply the number of matches.\footnote{The BioNetGen language follows the number-of-reactions convention for rate calculation, whereas the Kappa language follows the number-of-matches convention.}  As a result of this choice of convention, pathological cases requiring explicit accommodation in either the simulation engine or rule interpreter may arise in dissociation reactions when asymmetric patterns match symmetric molecules (see Section \ref{sec:dissoc}).

\subsection{Reaction path degeneracy}\label{sec:rpd}
\begin{figure}[t]
\centering
\includegraphics[width=0.65\textwidth]{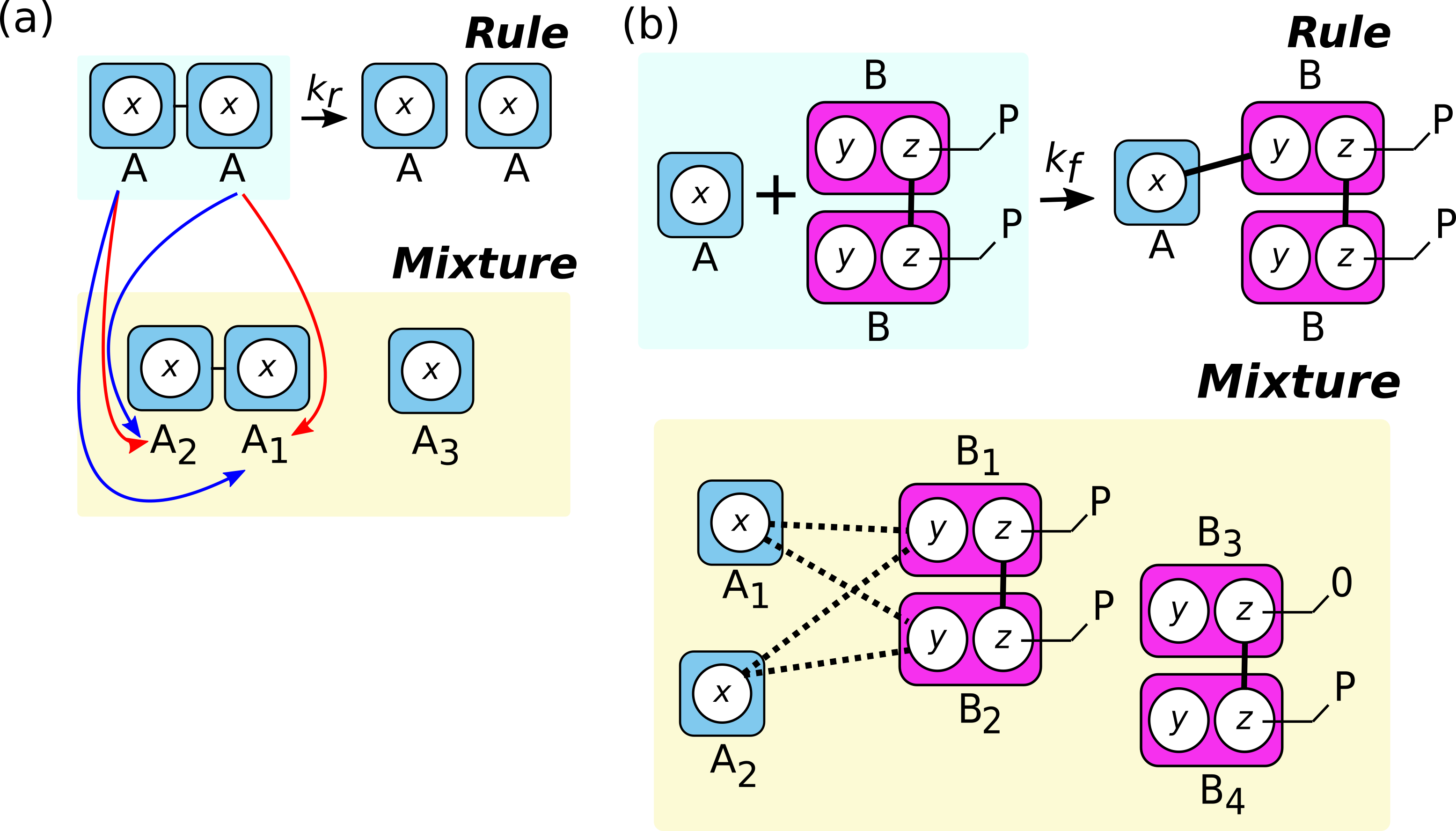}
\caption{Symmetry and reaction path degeneracy influence rule rate calcuation.  \textbf{a} Symmetry in a pattern results in multiple matches (red and blue arrows) from the pattern to a particular set of molecules.  Correctly computing the rate of a rule (given certain assumptions outlined in Section \ref{sec:permute}) requires that the rate is divided by the number of symmetries present in the rule's patterns.  \textbf{b} Rules where a pattern has multiple sites or pattern molecules that can be the target of the transformation defined by a rule have reaction path degeneracy.  This can be seen in the simulation mixture illustrated at the bottom of the panel, where there are four possible binding events (denoted by the broken lines) implied by the rule illustrated at the top of the panel.  Correctly calculating the rate for the rule requires multiplying the match-based rate by the number of possible reaction paths present among the rule's patterns (2 $y$ sites competent for binding site $x$ on $A$) and also accommodating other symmetries effects if necessary (dividing by 2 due to the symmetric $B$--$B$ dimer pattern).}
\label{fig:complex}
\end{figure}

A pattern may have identical sites to which a transformation defined by a rule (e.g., modification of a site's internal state or formation/dissolution of a bond) may apply equally.  This is termed reaction path degeneracy \citep{iupac2009goldbook}, referring to multiple equivalent ways for a reaction to occur with respect to some particular biochemical species.  To calculate rule rates such that the rate constant refers to the reaction applied to an individual site (or pair of sites in the case of bond formation), the number of matches in the pattern's match list is multiplied by the number of equivalent sites in the pattern that are competent to be modified by the transformation defined by a rule (Fig. \ref{fig:complex}b).  The multiplicative factor used to adjust the rate in cases of reaction path degeneracy may also be termed a statistical factor.  During the rule application process, a random site is selected to undergo the transformation defined by a rule with uniform probability.  In the case of Fig. \ref{fig:complex}b, the rate is $4\cdot k_f$: 
\begin{displaymath}
\left(\dfrac{2_1\cdot 2_2 \cdot 2_4}{2_3}\right)\cdot k_f = 4 \cdot k_f
\end{displaymath}
where the subscripts denote:
\begin{enumerate}
\item the number of matches of the $A$ pattern
\item the number of matches of the $B$ dimer pattern
\item the correction for symmetry in the $B$ dimer pattern
\item the number of equivalent reaction paths\footnote{The reaction path degeneracy is 2 because binding to either $y$ site on the $B$ dimer pattern results in the same species instance.}
\end{enumerate}
Note that reaction path degeneracy often coincides with the presence of pattern-preserving site permutations or symmetry, and in these cases, the rate must be further adjusted as described in Section \ref{sec:permute}.  

Regardless of the framework's chosen convention, the user must be aware of how the matching procedure may influence a rule's rate to correctly specify the rate constants in the rate laws associated with the rules (Section \ref{sec:discuss}). 

\subsection{Molecularity}\label{sec:molec}
\begin{figure}[t]
    \centering
  \includegraphics[width=0.52\textwidth]{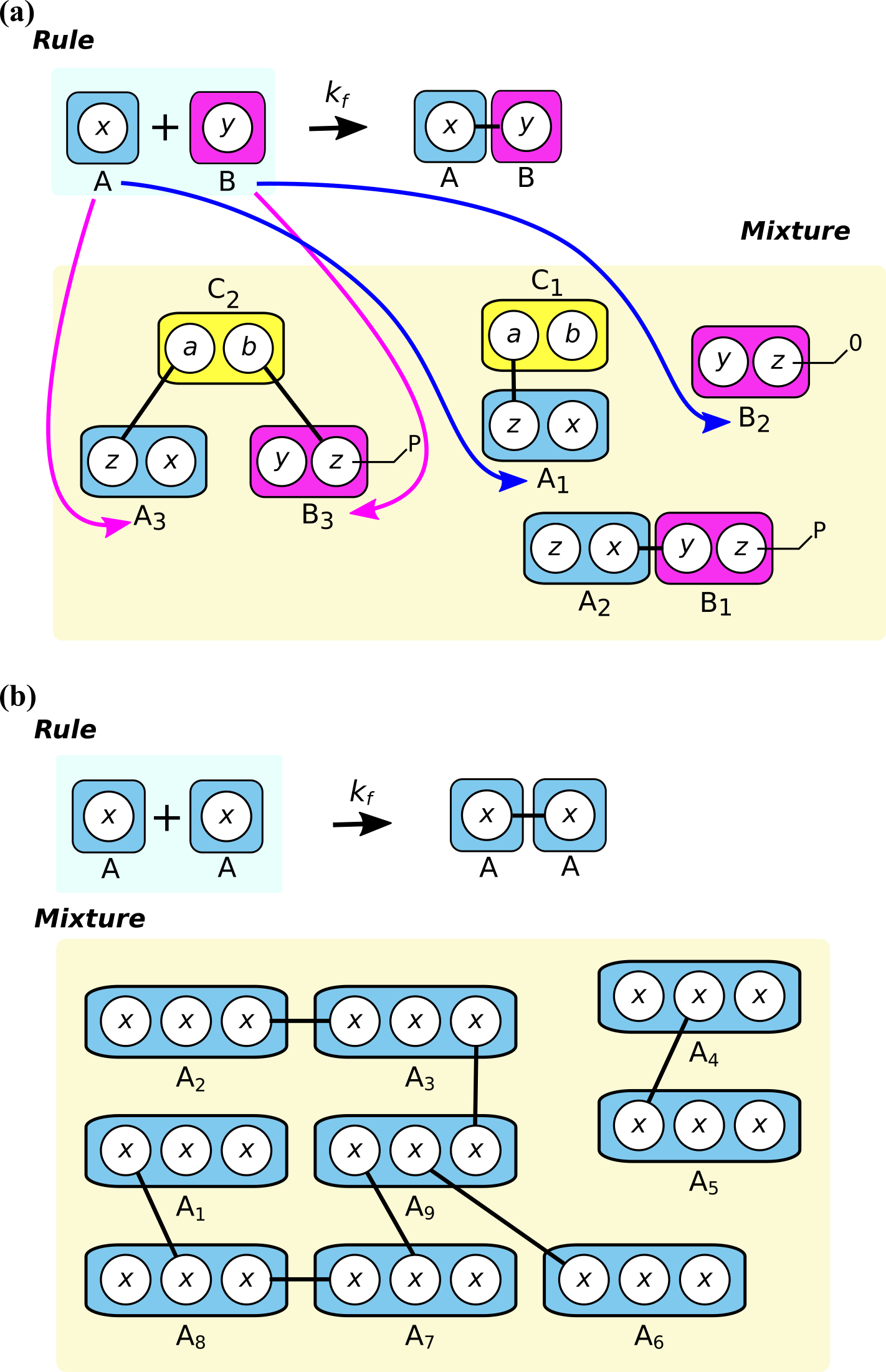}
\caption{Null events arising from molecularity constraints. \textbf{a} Selecting the match corresponding to the magenta arrows would fail to satisfy molecularity, resulting in a null event, whereas selecting the match corresponding to the blue arrows does satisfy the molecularity constraint of the rule. \textbf{b} Example rule and mixture in which over half of all potential reaction events would be null events, making for a very inefficient simulation.}
\label{fig:4}       
\end{figure}

    An additional consideration is the molecularity of rules.  When species composed of more than one molecule exist in a simulation mixture, it is possible for multiple patterns in a single rule to match the same species instance.  For example, suppose we have a bond formation rule stating that pattern molecule $A$ with free site $x$ may bind to pattern molecule $B$ with free site $y$, and the mixture contains a trimeric complex, in which molecules $A$ and $B$ satisfy the patterns' constraints, but $A$ and $B$ are also each bound to a molecule $C$ via sites unspecified in the bond formation rule (Fig. \ref{fig:4}a).  When this rule is sampled, the algorithm might randomly choose matches corresponding to the molecules $A$ and $B$ that are members of the trimeric complex.  The rule states that $A$ should bind to $B$, forming a cyclic structure, although this is not obvious from the rule's reactant pattern.  Furthermore, the rate of cyclization (a unimolecular reaction) will not be equivalent to the rate of bimolecular association, and so a distinction should be made between intermolecular and intramolecular bond formation.

To distinguish between unimolecular and bimolecular reactions, the simulator must perform a check to confirm that matches for a bimolecular rule are members of distinct species instances and not members of the same species instance. This can be done by traversing the matched molecules' species instances as is done when updating the system after a reaction event; the task can also be accomplished by tracking species instances with unique identifiers (see Section \ref{sec:null}).  

\subsection{Locality}\label{sec:local}
Rules may include negative application conditions (i.e., constraints on when they may be applied).  Such conditions are especially relevant when considering the molecularity of rules in the previous section (i.e., the bimolecular vs. unimolecular bond formation).  It is sometimes possible to ignore molecularity constraints in the determination of matches between patterns and molecules (e.g., when intramolecular binding is not possible).  Consider the rule set from Fig \ref{fig:1}c and specifically the binding rule.  Regardless of the sequence or number of reaction events in a mixture initially composed of $A$ and $B$ monomers, we know that the patterns in the first rule will only match monomeric $A$ and $B$ molecules without any additional enforced constraint that the transformation defined by a rule be bimolecular (i.e., the rule will never result in an intramolecular binding event).  When molecularity does not need to be checked, a rule's application is local, meaning that no information outside the molecules involved in the patterns' matches is required to correctly sample matches, apply rules, and update the system \citep{Harmer2010}.

A strongly contrasting phenomenon exists when allowing patterns to include implicit bonds, meaning that a rule-based modeling framework can enforce connectivity between molecules without specifying how the molecules are connected.  If a rule contains implicit bonds, prediction of which rule rates to update cannot be done until after application of the rule to a specific species instance.  We accommodate the presence of implicit bonds in our description of the update scheme (Section \ref{sec:distinct}) by requiring new species to be checked against \emph{all} rule patterns for new matches.

However, if no rules in a model involve implicit bonds, then more efficient updating schemes can be realized. One example is a structure that can be computed directly from the set of rules, and relates how the application of one rule may increase or decrease the rate of another rule\footnote{This is termed the influence map in the Kappa language and associated simulation engines.} similar to a dependency graph for traditional SSA implementations \citep{Danos2007scalable,Danos2009}. For special case models where molecularity in rules does not need to be checked, the system can even be updated without the need to traverse the species instances involved in a particular reaction \citep{Danos2007scalable}.

\subsection{Null events}\label{sec:null}
In the previously described problem of constraining molecularity (Section \ref{sec:molec}), there may be occurrences in which the rates are overestimated, because some choices of matches lead to invalid reactions (Fig. \ref{fig:4}a). To correct for these overestimates, some network-free simulation approaches allow null events.

If the sampled molecules fail to satisfy the molecularity constraints in the sampled rule (Fig. \ref{fig:4}a), then the potential reaction event is rejected, time is updated, and the algorithm proceeds to the next iteration with an identical system configuration \citep{Yang2008kinetic}.  The time update without a corresponding system update corrects for the overestimated rate; the null event uses up the excess rate resulting from the invalid combination of sampled matches.

This type of null event can become a serious computational inefficiency in models that tend to form large aggregates.  Consider a bimolecular association rule.  If most of the molecules in a system are part of the same, large aggregate (Fig. \ref{fig:4}b), it is possible that the majority of iterations of the algorithm will choose two matches from within the same species instance.  However, the rule requires an intermolecular bond to form, yielding a correspondingly large number of null events.  Alternative implementations can be realized that avoid such issues (see Section \ref{sec:discuss}).

A related type of null event may arise when sampling matches from a match list.  Consider a homodimerization rule whose two patterns are identical.  Upon sampling this rule and a match for the first pattern, it is possible that the same match may be sampled for the second pattern.  If the matches overlap (i.e., they involve the same molecules' sites) this results in a null event, termed a clash, where time is similarly updated and the system remains the same \citep{Danos2007scalable}.  For example, Fig. \ref{fig:4}b shows a dimerization rule.  In this case both reactant patterns in the rule can match any molecule in the mixture.  A clash could occur in this system when, during match sampling (Step 3 in Section \ref{sec:genericAlgo}), both matches involve the same molecule in the mixture (e.g., the selected matches for both reactant patterns point to $A_6$).

\section{Accommodating rule conventions}\label{sec:conventions}

\subsection{Dissociation pathologies}\label{sec:dissoc}
Briefly mentioned in Section \ref{sec:permute} was the possibility of pathological cases for rate calculations of dissociation rules. This occurs when a pattern that has no pattern-preserving site permutations (i.e., symmetries, see Section \ref{sec:permute}) matches the same set of molecules multiple times (Fig. \ref{fig:pathology}a). This stems from our convention that a rule's rate should be proportional to the number of distinct reactions that can occur as opposed to the number of matches of the pattern in the mixture. Such pathologies do not occur when a rule's rate is proportional to the number of matches (Fig. \ref{fig:pathology}c); however, assuming a number-of-matches convention detracts from the physical meaning of the rule's rate constant (generally considered to describe the rate of bond dissolution).  The choice of which convention to follow is a matter of design; the important point is that the user understand which convention is in use to avoid writing rules that have unexpected consequences.

\afterpage{
\begin{figure}[t]
\centering
\includegraphics[width=0.6\textwidth]{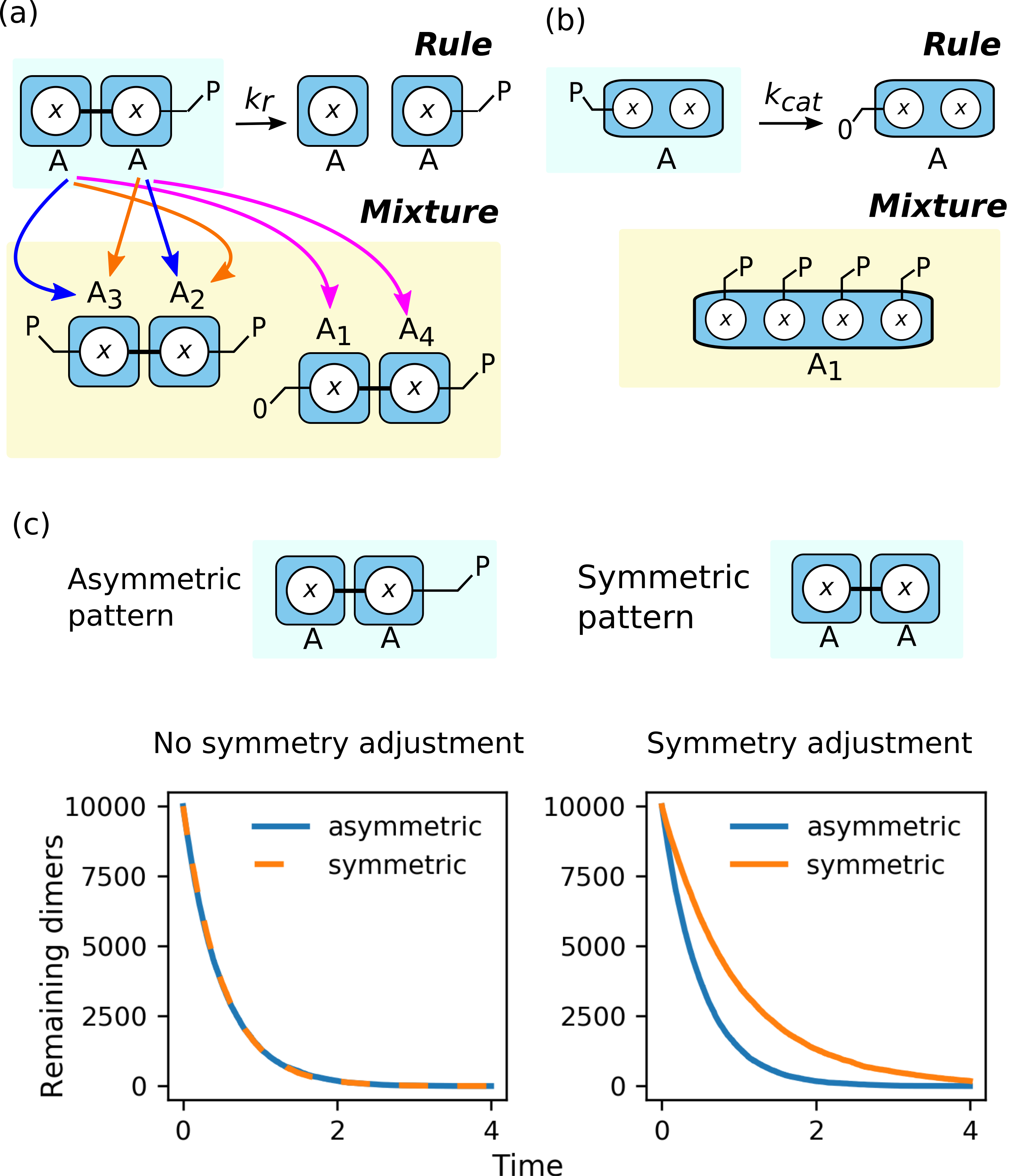}
\caption{ Certain pathological behaviors may arise due to language conventions. \textbf{a} Dimer dissociation example, where the pattern matches one potential reactant twice and another potential reactant once. If dissociation rate constants refer to the rate of a single bond breaking, then inconsistencies in rate calculation may arise when asymmetric patterns match symmetric molecules. \textbf{b} Dephosphorylation in the presence of multiple sites.  The rate of this rule depends on interpretation.  If each site $x$ should be dephosphorylated independently with rate $k_{cat}$ and the only other constraint is at least one unbound site $x$, then a hierarchical matching scheme needs to be implemented (Section \ref{sec:samesites}). \textbf{c} Curves for dimer dissociation (as described in panel a) resulting from asymmetric patterns and symmetric patterns applied to the same system of symmetric species ($A$--$A$ homodimers with all sites in the `P' state). When the simulator automatically adjusts for pattern symmetry (Section \ref{sec:permute}), there is a discrepancy in the rate calculation that is observed in distinct dissociation curves (right), compared to a paradigm in which the rate constants are applied to the number of matches directly (left).}
\label{fig:pathology}
\end{figure}
\clearpage
}

In our presented framework, these cases must be distinctly considered. One possible approach would be to match bonds instead of molecules for dissociation rules. This would involve an initial check of the rule set to determine whether this pathology may arise. The check would perform a bidirectional site specificity check\footnote{For each site in one of the pattern molecules, the corresponding site in the other pattern molecule must either be equivalent or less specific and consistent.} for the two connected pattern molecules whose bond would be broken by the rule. Patterns that satisfy this check may produce multiple molecular matches to the same set of molecules in the mixture. Upon identification of these patterns, one may then use appropriate data structures to track bond matches instead of molecule matches.

\subsection{Identical site pathologies}\label{sec:samesites}
In some rule-based modeling frameworks, individual molecules are allowed to have identical sites.  In cases involving molecules with identical sites, further care must be taken to correctly calculate rule rates, which we do not explicitly discuss in our specification of a simulation engine.  Consider the rule and mixture visualized in Fig. \ref{fig:pathology}b.  If the algorithm generates all possible matches between the pattern sites and the single molecule in the mixture, $A_1$, there would be 12 matches (the 2-permutations of the set of 4 sites).  However if the prescribed rate constant refers to the rate at which a site $x$ in molecule $A$ is dephosphorylated, the rule's rate would be incorrectly calculated as $12\cdot 1\cdot k_{cat}$ instead of $4\cdot 1\cdot k_{cat}$.  

Additional machinery beyond what we describe here is needed to accommodate such a problem.  If the language interprets the rule in Fig. \ref{fig:pathology}b as having a rate of reaction equal to $4\cdot 1\cdot k_{cat}$, then a possible solution would be to introduce an additional matching procedure that distinguishes between the sites in the pattern or pattern molecule based on how the sites participate in the transformation defined by a rule.  Sites that are modified by the transformation defined by a rule would be the primary matching sites (e.g., the phosphorylated site $x$ in Fig. \ref{fig:pathology}b).  The other sites that provide context, but are unmodified by reaction would not contribute to additional matches (e.g., the partially specified site $x$ in Fig. \ref{fig:pathology}b).  Of course if the modeler intends molecule $A_1$ to be dephosphorylated proportional to the number of matches, then the original formulation would be desired, with rate $12\cdot 1\cdot k_{cat}$.  Being able to specify both types of rules would require specific language features, or some sort of annotation so that the simulation engine correctly interprets how a rule's rate should be calculated in the presence of identical sites.

In general, care must be taken when designing network-free simulation algorithms for specific modeling languages.  Features or conventions present in the modeling language, such as those described in this section, can result in language-specific behaviors.  This is especially evident when attempting to translate a model into a different rule-based modeling language \citep{Suderman2017truml}.  Documentation of pathological cases (both for the modeling language and the simulation engine) as well as the semantics of the rule-based modeling language are essential for accurate and consistent modeling and simulation.

\section{Discussion} \label{sec:discuss}

Network-free simulation has now been around for a decade, and its continued use in dynamical systems biology research is strong evidence in favor of the utility of the methodology.  Our aim here was to provide a foundation for developing a network-free, kinetic Monte Carlo simulation engine for rule-based models.  Beyond the more accessible high-level description of how Gillespie's direct method is generalized for rule-based models, we proceeded to define a basic framework and document a number of nontrivial implementation details required for constructing a network-free simulation engine.  While our approach covers a wide range of use cases, its explicit implementation is not general and requires extension to accommodate the edge cases described in Sections \ref{sec:dissoc}, \ref{sec:samesites} and perhaps others as well.  

As the title states, the network-free methodology described here is a generalization of Gillespie's direct method.  However, its impact is broader than simply a new approach to capture stochastic fluctuations in biochemical systems with small population sizes.  Rule-based modeling, coupled with network-free simulation, enables modelers to define classes of reactions based on limited interaction information (i.e., omitting molecular context that is either irrelevant or not known to alter the rate of reaction for a particular molecular moiety), precluding the need to enumerate all species and reactions for a particular interaction network.  Indeed, generating the entire reaction network is often impossible or not computationally feasible.  Explicitly accommodating combinatorial complexity enables more detailed and more precise investigation of system dynamics without unjustified simplification of models \citep{Suderman2013,Deeds2012,Faeder2005}.  Accommodating such complexity is especially relevant for characterizing systems involving biopolymers \citep{Kohler2014,Aitken2013} or systems that can undergo a phase transition to a gel state \citep{Goldstein1984}.  Clearly, these studies and others that use network-free simulation engines are not at all concerned with stochastic effects, but with the ability to model and simulate these biochemical systems that contain a high degree of combinatorial complexity \citep{Stites2015,Creamer2012}.  Network-free simulation algorithms provide the only available, general framework for exactly simulating the dynamics of such systems.  

Frequent use of existing software suites that have network-free simulation capabilities has led to the discovery of software bugs and inefficiencies as well as the desire for new features.  Indeed, the field of network-free simulation algorithm development is not as mature as a that of the SSA field.  Ongoing work seeks to improve the performance of network-free simulation \citep{Boutillier2017} and there is much work to do.  Upon considering the developmental trajectory of the traditional SSA based on Gillespie's direct method, we anticipate that increasing applications of rule-based modeling and network-free simulation will both drive innovation in network-free simulation algorithms and motivate hardening of the software.  

As an example of such innovation, some of the complexities discussed in Section \ref{sec:complexity} can be eliminated at the cost of additional computational overhead with an alternative sampling implementation.  Driven by the need to simulate systems that result in a high proportion of null events, species instances themselves could be explicitly tracked throughout the simulation and matches would be associations between patterns and unique sets of molecules as outlined in \citep{Colvin2010RuleMonkey}.  While this comes at significant computational cost, it is demonstrably useful in models that can produce gels or other large polymers, because it eliminates the need for null events (i.e., a rejection-free algorithm as opposed to the rejection algorithm we describe here) \citep{Yang2011}.  One useful extension of existing network-free simulation packages might be to implement adaptive algorithm selection that determines on-the-fly whether rejection or rejection-free methods are more appropriate for simulating a particular model's dynamics (e.g., whether or not the mixture contains large polymers).  Another straightforward extension of network-free simulation that is becoming increasingly relevant is to consider spatial as well as temporal dynamics \citep{Klann2012}.  As available computing power increases, performing increasingly complex spatial simulations will become feasible.  Integrating spatial simulation engines with network-free algorithms are already beginning to emerge \citep{Kochanczyk2017,Sorokina2013,Tapia2016}.  Ultimately, we expect that network-free simulation will play an increasingly prominent role in the modeling of complex biological dynamics.  

\section{Acknowledgements}
We thank Jim Faeder for providing helpful feedback on the manuscript.  This work was supported by the National Institute of General Medical Sciences (NIGMS) and the National Cancer Institute (NCI) of the National Institutes of Health (NIH) through grants R01GM111510, P50GM085273, and R01CA197398; by the U.S. Department of Energy (DOE) through contract DE-AC52-06NA25396; and by the Joint Design of Advanced Computing Solutions for Cancer (JDACS4C) program established by DOE and NCI/NIH. Additionally, RS, YTL and SF gratefully acknowledge support from the Center for Nonlinear Studies (CNLS), which is funded by the Laboratory Directed Research and Development (LDRD) program at Los Alamos National Laboratory.

\appendix

\section{PDGFR Activation Model}\label{sec:pdgfr}
\begin{lstlisting}[language=bngl,caption={A model for platelet-derived growth factor receptor (PDGFR) activation written in the BioNetGen language. PDGFR can dimerize if at least one PDGFR is bound to ligand.  Dimerized PDGFR can then undergo autophosphorylation of 10 distinct tyrosine residues, and phosphorylation (and dephosphorylation) on each residue occurs independently.  Here, all rate constants are set to 1, and the `\textbackslash' character denotes line continuation for clarity.  Actual simulation of the model with NFsim (see the `actions' block) requires consolidating the continued lines in the `molecule types' and `seed species' blocks.},label={lst:pdgfr}]
begin model
begin molecule types
  PDGFR(lig,pdgfr,y1~0~P,y2~0~P,y3~0~P,y4~0~P,y5~0~P,\
     y6~0~P,y7~0~P,y8~0~P,y9~0~P,y10~0~P)
  Lig(pdgfr)
end molecule types
begin seed species
  PDGFR(lig,pdgfr,y1~0,y2~0,y3~0,y4~0,y5~0,y6~0,\
        y7~0,y8~0,y9~0,y10~0) 1000
  Lig(pdgfr) 10000
end seed species
begin observables
Species PDGFR_dimers PDGFR(pdgfr!1).PDGFR(pdgfr!1)
Molecules phospho_PDGFR  PDGFR(y1~0) PDGFR(y2~0) PDGFR(y3~0)\
                         PDGFR(y4~0) PDGFR(y5~0) PDGFR(y6~0)\
                         PDGFR(y7~0) PDGFR(y8~0) PDGFR(y9~0)\
                         PDGFR(y10~0)
end observables
begin reaction rules
  PDGFR(lig) + Lig(pdgfr) <-> PDGFR(lig!1).Lig(pdgfr!1) 1, 1
  PDGFR(lig!+,pdgfr) + PDGFR(pdgfr) <-> \
    PDGFR(lig!+,pdgfr!1).PDGFR(pdgfr!1) 1, 1
  PDGFR(pdgfr!+,y1~0) -> PDGFR(pdgfr!+,y1~P) 1
  PDGFR(pdgfr!+,y2~0) -> PDGFR(pdgfr!+,y2~P) 1
  PDGFR(pdgfr!+,y3~0) -> PDGFR(pdgfr!+,y3~P) 1
  PDGFR(pdgfr!+,y4~0) -> PDGFR(pdgfr!+,y4~P) 1
  PDGFR(pdgfr!+,y5~0) -> PDGFR(pdgfr!+,y5~P) 1
  PDGFR(pdgfr!+,y6~0) -> PDGFR(pdgfr!+,y6~P) 1
  PDGFR(pdgfr!+,y7~0) -> PDGFR(pdgfr!+,y7~P) 1
  PDGFR(pdgfr!+,y8~0) -> PDGFR(pdgfr!+,y8~P) 1
  PDGFR(pdgfr!+,y9~0) -> PDGFR(pdgfr!+,y9~P) 1
  PDGFR(pdgfr!+,y10~0) -> PDGFR(pdgfr!+,y10~P) 1
  PDGFR(y1~P) -> PDGFR(y1~0) 1
  PDGFR(y2~P) -> PDGFR(y2~0) 1
  PDGFR(y3~P) -> PDGFR(y3~0) 1
  PDGFR(y4~P) -> PDGFR(y4~0) 1
  PDGFR(y5~P) -> PDGFR(y5~0) 1
  PDGFR(y6~P) -> PDGFR(y6~0) 1
  PDGFR(y7~P) -> PDGFR(y7~0) 1
  PDGFR(y8~P) -> PDGFR(y8~0) 1
  PDGFR(y9~P) -> PDGFR(y9~0) 1
  PDGFR(y10~P) -> PDGFR(y10~0) 1
end reaction rules
end model
begin actions
simulate({method=>"nf",t_start=>0,t_end=>100,n_steps=>100})
end actions
\end{lstlisting}

\end{document}